\documentclass[aps,eqsecnum,nofootinbib,superscriptaddress,showpacs,twocolumn]{revtex4}
\def\be{\begin{eqnarray}}
\def\ee{\end{eqnarray}}
\usepackage[dvipdfmx]{graphicx}
\usepackage{amsfonts,bm}
\usepackage{amsmath,amsfonts,amssymb,bm}
\usepackage[dvips]{color}
\def\p{\partial}

\newcommand{\Exp}[1]{\left\langle~#1~\right\rangle}

\begin{document}
\title{Achronal ANEC, Weak Cosmic Censorship, and AdS/CFT duality}


\author{Akihiro {\sc Ishibashi}}\email[]{akihiro@phys.kindai.ac.jp}
\affiliation{%
{\it Department of Physics and Research Institute for Science and Technology, Kindai University, Higashi-Osaka 577-8502, JAPAN
}}

\author{Kengo {\sc Maeda}}\email[]{maeda302@sic.shibaura-it.ac.jp}
\affiliation{%
{\it Faculty of Engineering,
Shibaura Institute of Technology, Saitama 330-8570, JAPAN}}

\author{Eric {\sc Mefford}}\email[]{eric.mefford@polytechnique.edu}
\affiliation{{\it Centre de Physique Th\'eorique, \'Ecole Polytechnique, CNRS, Route de Saclay, 91128 Palaiseau, FRANCE}}

\begin{abstract}
We examine the achronal averaged null energy condition~(ANEC) for a class of conformal field theories~(CFT) at 
strong coupling in curved spacetime. By applying the AdS/CFT duality, we find holographic models which 
violate the achronal ANEC for $3+1$ and $4+1$-dimensional boundary theories. In our model, the bulk spacetime 
is an asymptotically AdS vacuum bubble solution with neither causality violation nor singularities. 
The conformal boundary of our bubble solution is asymptotically flat and is causally proper in the sense that 
 a ``fastest null geodesics" connecting any two points on the boundary must lie entirely on the 
boundary. We show that conversely, if the spacetime fails to have this causally proper nature, then there must be 
a naked singularity in the bulk.   
\end{abstract}

\maketitle

\section{Introduction}
The classical null energy condition (NEC) $T_{\mu\nu}l^\mu l^\nu\ge 0$ for all null vectors $l$ at all points is the key condition for the proof of the singularity theorems, 
topological censorship theorem, and other important theorems in classical general relativity. Although it is satisfied for typical classical matter fields,  
the NEC can be violated when one considers quantum field effects. 
For example, when one spatial dimension is compactified, $\Exp{T_{\mu\nu}}l^\mu l^\nu$ becomes negative 
along the null geodesic of the closed $S^1$ circle, where $\Exp{T_{\mu\nu}}$ is the vacuum expectation value of the stress-energy tensor. In general, any locally formulated energy conditions can be violated by quantum field effects. 

The averaged null energy condition~(ANEC) is an alternative condition which is non-locally formulated and states that 
\begin{align}
\label{ANEC} 
\int^{\infty}_{-\infty}\Exp{T_{\mu\nu}} l^\mu l^\nu d\lambda\ge 0 \,,  
\end{align}
for every complete null geodesic with tangent vector $l^\mu$, where $\lambda$ is the affine parameter. 
It has been shown that the ANEC is satisfied for some cases, e.g., minimally coupled scalar fields in 
Minkowski spacetime~\cite{Klinkhammer1990, WaldYurtsever1991} in $4$-dimensions and in curved 
spacetime~\cite{Yurtsever1990, WaldYurtsever1991} in 2-dimensions. 
For further examples in which the ANEC holds, see Refs.~\cite{Ford_Roman1995, Yurtsever1995}. 
However, it has been shown that for a conformally coupled scalar field, the ANEC can be violated 
for any {\it chronal} null geodesics in Schwarzschild spacetime~\cite{Visser1996}. This example of ANEC violation has led Graham 
and Olum~\cite{GrahamOlum} to propose the {\it achronal} ANEC, which states that the ANEC should hold 
for every complete achronal null geodesic but not necessarily on chronal null geodesics. 
Here, an achronal null geodesic refers to a null geodesic curve on which no two points can be connected 
by a timelike curve. A complete achronal null geodesic is also called a null line. 
Further studies, however, revealed cases in which the achronal ANEC can also be violated~\cite{Visser1995,UrbanOlum}. 
This fact suggests the possibility of, e.g., the formation of exotic objects such as wormholes, 
since the achronal ANEC is crucial for the proof of singularity theorems and topological censorship.  
  
It is interesting to study whether the achronal ANEC is violated in a strongly coupled field theory in the framework of the AdS/CFT duality. 
In this context it was recently shown in~\cite{KellyWall2014} that the achronal ANEC holds for a class of conformal field theories 
in the boundary Minkowski spacetime. 
This is consistent with the numerical verification of the achronal ANEC for colliding planar shock wave solutions~\cite{derSchee2014}. 

Applying the AdS/CFT duality, an example of NEC violation in curved space was recently found in~\cite{IMM2018}, 
in which the gravity dual is a vacuum AdS black hole solution, and the boundary spacetime describes 
a wormhole geometry which connects two asymptotically flat universes. 
In this example, as the bulk solution asymptotically approaches the planar Schwarzschild-AdS solution, the corresponding boundary thermal states render  $\Exp{T_{\mu\nu}} l^\mu l^\nu$ strictly positive in the asymptotic region of the boundary spacetime. Therefore even though the NEC is locally violated near the wormhole throat, the achronal ANEC is kept preserved. This suggests that the achronal ANEC may always hold in a thermal state with asymptotically flat boundary spacetime. 

In this paper, we further examine the achronal ANEC for a class of strongly coupled field theories in asymptotically flat curved spacetimes 
in the framework of the AdS/CFT duality. 
If and when there exists a timelike curve in the bulk that connects two points on a boundary achronal null geodesic, one can say that the corresponding boundary theory admits an acausal signal. We shall reveal some possible relations between the achronal ANEC, weak cosmic censorship, and 
acausal propagation of signals, provide some examples of an achronal ANEC violation without acausal signals and finally discuss what happens 
when there are acausal signals.

We first give examples of a violation of the achronal ANEC in $d=4$ and $5$ boundary spacetimes with respectively, $d+1=5$ and $6$ bulk solutions of the vacuum Einstein equations with a negative cosmological constant. 
Our strategy is similar to Ref.~\cite{UrbanOlum}. In the case of $d=4$, we start with a vacuum bubble AdS solution, also called the AdS soliton, 
as our 5-dimensional bulk spacetime. We show that a certain choice of conformal factor for a conformally flat, $d=4$ boundary spacetime 
induces a gravitational conformal anomaly~\cite{HaroSkenderis}, which leads to the violation of the achronal ANEC. 
For $d=5$ spacetimes, there is no gravitational anomaly. We construct a regular $6$-dimensional bulk  
vacuum bubble solution with a curved $d=5$ boundary spacetime. 
The resulting boundary stress-energy tensor shows that the NEC on the boundary is violated but the ANEC on the boundary 
still holds. Then, by choosing a suitable conformal factor, we show that the achronal ANEC in the conformally transformed 
system can be violated.  

It is worth mentioning that both of our two examples are ``causally proper" in the sense that 
there is no bulk timelike curve that could connect any two points on each achronal null geodesic on the boundary, 
implying a ``fastest null geodesics" connecting any two points on the boundary must lie entirely on the boundary. 
In addition, there is no pathological behavior in the boundary spacetime: it is in fact geodesically complete and asymptotically flat. 
It is then interesting to consider what would possibly happen if, on the other hand, the causally proper nature 
is not satisfied. 
We will show by using the Gao-Wald theorem \cite{GaoWald} that if the geometry under consideration fails to be causally proper 
while preserving the NEC in the bulk, then there must appear a naked singularity; the weak cosmic censorship must fail.

In the next section we consider the achronal ANEC in $d=4$ boundary theory. In Sec.~\ref{sec:III}, we construct a $6$-dimensional perturbed 
bubble solution which leads to the violation of the ANEC. A proposition which connects the bulk cosmic censorship and the 
causally proper nature is presented in Sec.~\ref{sec:IV}. 
Sec.~\ref{sec:V} is devoted to summary and discussions.

\section{Violation of achronal ANEC in even dimensions}
\label{sec:II}
In this section, we explore the ANEC in a $5$-dimensional holographic model. 
In $d$-dimensional boundary theory with $d$ even, the bulk metric is expressed in the Fefferman-Graham coordinate system:  
\begin{align}
\label{Fefferman-Graham}
ds^2=\frac{1}{z^2}\left[dz^2+\Bigl(\sum_{n=0}^\infty g_{(n)\mu\nu}z^n+ z^d\ln z^2 h_{\mu \nu} \Bigr) dx^\mu dx^\nu\right]
\end{align} 
with the boundary metric $ds_\p^2=g_{(0)\mu\nu}dx^\mu dx^\nu$ located at $z=0$~\cite{HaroSkenderis}. 
Here, the logarithmic term only appears for even dimensions, and $g_{(2k+1)\mu\nu}=0$ for any integer $k$ satisfying 
$0\le 2k+1<d$. 
The holographic stress-energy tensor $\Exp{T_{\mu\nu}}$ includes conformal anomalies, and it is given by  
\begin{align}
\label{even-stress-energy}
& \Exp{T_{\mu\nu}}=g_{(4)\mu\nu}-\frac{1}{8}g_{(0)\mu\nu}
\left\{(g^\alpha_{(2)\alpha})^2-g^\alpha_{(2)\beta}g^\beta_{(2)\alpha}  \right\} \nonumber \\
&\quad\quad\quad\quad-\frac{1}{2}g_{(2)\mu\alpha}g^\alpha_{(2)\nu}+\frac{1}{4}g_{(2)\mu\nu}g^\alpha_{(2)\alpha},  
\end{align}
where the gravitational constant $G_4$ is set to be $4\pi G_4=1$ and the index is raised and lowered by 
the boundary metric $g_{(0)\mu\nu}$. 
The coefficient $g_{(2)}$ is induced by the Ricci tensor on the boundary metric $ds_\p^2$~\cite{HaroSkenderis}, so we consider a conformal 
transformation of the boundary metric $d\hat{s}_\p^2=a(x^\mu)^2ds_\p^2$ and investigate whether the achronal ANEC 
can be violated for the boundary conformal field theory. 
Note that in a conformally flat spacetime, any null geodesics are achronal. 
We are interested in the boundary metric where the null geodesics are complete in both future and past directions 
and the l.~h.~s. of Eq.~(\ref{ANEC}) converges. 
We consider a conformal factor $a(x^\mu)$ satisfying the following conditions; \\
\\
{\it Condition~A}
\begin{enumerate}
\item  $a(\lambda)$ is everywhere regular (at least twice
differentiable) and positive-definite. 
\item At $\lambda \rightarrow \pm \infty$, $a(\lambda)$ approaches some finite positive constant values.    
\end{enumerate}
where $\lambda$ is the affine parameter of the null geodesic.  

\subsection{Achronal ANEC for a simple case}
We start with a 5-dimensional vacuum bubble solution with the metric,  
\begin{align}
\label{bubble-even}
ds^2=\left(r^2-\frac{r_0^4}{r^2} \right)d\chi^2+\frac{dr^2}{\left(r^2-\frac{r_0^4}{r^2} \right)}+r^2(dx^2+dy^2-dt^2),  
\end{align}
where $\chi\in [0, \,\pi/r_0]$, and the (conformal) boundary metric is $ds_\p^2=-dt^2+dx^2+dy^2+d\chi^2$. 
A good place to start is to consider first the case where $a$ depends on $x$ only. As done in the cosmological case~\cite{Siopsis2009},   
one needs to make a change of coordinates $(r,\,x)\to (z,\,\rho)$ to bring (\ref{bubble-even}) to the Fefferman-Graham coordinate (\ref{Fefferman-Graham}) 
with the boundary metric $d\hat{s}_\p^2=a(x)^2ds_\p^2$. 

Introducing new coordinates $z$ and $\rho$ as 
\begin{align}
& \frac{1}{r}=z\left(\frac{1}{a(x)}+\alpha_1(x)z^2+\alpha_2(x)z^4+\cdots     \right), \nonumber \\
 & x(\rho,\,z)=\rho+\beta_1(\rho)z^2+\beta_2(\rho)z^4+\cdots, \nonumber \\
 &a(x)=a(\rho)+a'(\rho)\beta_1(\rho)z^2+\Bigl(a'(\rho)\beta_2(\rho)\nonumber\\
& \quad\quad\quad+\frac{1}{2}\beta_1^2(\rho)a''(\rho)\Bigr)z^4+\cdots, 
\end{align}
the bubble metric~(\ref{bubble-even}) is reduced to the Fefferman-Graham metric~(\ref{Fefferman-Graham}) 
under the conditions 
\begin{align}
& \alpha_1(\rho)=-a(\rho)\beta_1^2(\rho),\;\; \alpha_2(\rho)=\frac{a'(\rho)^4-2r_0^4\,a(\rho)^4}{16a(\rho)^9}, \nonumber \\
& \beta_1(\rho)=\frac{a'(\rho)}{2a(\rho)^3}, \quad \beta_2(\rho)=-\frac{a'(\rho)^3}{8a(\rho)^7}, \cdots,  
\end{align} 
where $a'=\p_\rho a$. 
For our purpose, the other higher order coefficients are not needed, as we are only concerned with the derivation of 
Eq.~(\ref{even-stress-energy}). The coordinate transformation just corresponds to choosing a different foliation from 
the original bubble solution~(\ref{bubble-even}).  

Each coefficient in the Fefferman-Graham metric is given by 
\begin{align}
& g^{(0)}_{\mu\nu}dx^\mu dx^\nu=a(\rho)^2(-dt^2+d\rho^2+dy^2+d\chi^2), \nonumber \\
& g^{(2)}_{\mu\nu}dx^\mu dx^\nu=-\frac{a'(\rho)^2}{2a(\rho)^2}dt^2 \nonumber \\
&+\frac{2a(\rho)a''(\rho)-3a'(\rho)^2}{2a(\rho)^2}d\rho^2
+\frac{a'(\rho)^2}{2a(\rho)^2}(dy^2+d\chi^2), \nonumber \\
& g^{(4)}_{\mu\nu}dx^\mu dx^\nu=\frac{4r_0^4\,a(\rho)^4+a'(\rho)^4}{16a(\rho)^6}(-dt^2+dy^2) \nonumber \\
&+\frac{-12r_0^4\,a(\rho)^4+a'(\rho)^4}{16a(\rho)^6}d\chi^2 +d\rho^2\frac{1}{16a(\rho)^6}\times\nonumber\\
&\biggl\{9a'(\rho)^4+4a(\rho)[r_0^4\,a(\rho)^3-3a'(\rho)^2a''(\rho)+a(\rho)a''(\rho)^2]\biggr\}.
\end{align}
Substituting the above coefficients into Eq.~(\ref{even-stress-energy}), one obtains 
\begin{align}
& \Exp{T_{tt}}=-\frac{4r_0^4\,a(\rho)^4+5a'(\rho)^4-4a(\rho)a'(\rho)^2a''(\rho)}{16a(\rho)^6}, \nonumber \\
& \Exp{T_{\rho\rho}}=\frac{4r_0^4\,a(\rho)^4-3a'(\rho)^4}{16a(\rho)^6}, \nonumber \\
& \Exp{T_{yy}}=\frac{4r_0^4\,a(\rho)^4+5a'(\rho)^4-4a(\rho)a'(\rho)^2a''(\rho)}{16a(\rho)^6}, \nonumber \\
& \Exp{T_{\chi\chi}}=-\frac{12r_0^4\,a(\rho)^4-5a'(\rho)^4+4a(\rho)a'(\rho)^2a''(\rho)}{16a(\rho)^6}. 
\end{align}
Note that $\Exp{{T^\mu}_\mu}\neq 0$ unless $a(\rho)$ is constant. This is the effect of the conformal anomaly, which 
appears only for even dimensions. Now consider the null geodesic generator  
\begin{align}
\hat{l}=\frac{1}{a^2(x)}(\p_t+\p_x)
\end{align}
on the conformal boundary. Since the boundary metric is conformally flat, and the null geodesic orbit does not change for any 
conformal transformation, the null geodesic curve generated by $\hat{l}$ is achronal on the boundary theory.  

Under the condition~A, the l.~h.~s. of Eq.~(\ref{ANEC}) is evaluated as  
\begin{align}
& I=\int^\infty_{-\infty} T_{\mu\nu}\hat{l}^\mu\,\hat{l}^\nu\, d\lambda \nonumber \\
&=\int^\infty_{-\infty}\frac{a(\rho)a'(\rho)^2a''(\rho)-2a'(\rho)^4}{4a(\rho)^8}d\rho \nonumber \\
&=\int^\infty_{-\infty}\frac{a'(\rho)^4}{12a(\rho)^8}d\rho\ge 0, 
\end{align}
where we used condition~A to derive the last equality by integration by parts. 
This means that the achronal ANEC is satisfied for any conformal 
factor $a(\rho)$ satisfying the asymptotic boundary condition~A. 

\subsection{Violation of achronal ANEC for a generic scale factor $a$}
In this subsection we consider the case of generic scale factor $a(t,x,y)$, which depends on $t$, $x$, and $y$. 
Introducing new coordinates $\hat{x}^\mu~(\mu=0,1,2)$ and $z$ 
as  
\begin{align}
& \frac{1}{r}=z\left(\frac{1}{a(x)}+\alpha_1(x)z^2+\alpha_2(x)z^4+\cdots     \right), \nonumber \\
&  x^\mu(\hat{x},\,z)=\hat{x}^\mu+\beta_1^\mu(\hat{x})z^2+\beta_2^\mu(\hat{x})z^4+\cdots,  
\end{align}
we obtain the Fefferman-Graham metric~(\ref{Fefferman-Graham}) under the conditions  
\begin{align}
& \beta_1^\mu=\frac{\nabla^\mu a}{2a^3}, \qquad \beta_2^\mu=-\frac{\nabla^\mu a(\nabla a)^2}{8a^7}, \nonumber \\
& \alpha_1=-\frac{(\nabla a)^2}{4a^5}, 
\qquad \alpha_2=\frac{-2r_0^4\,a^4+((\nabla a)^2)^2}{16a^9}, \cdots, 
\end{align}
where $\nabla$ is the covariant derivative with respect to the metric $ds^2=-dt^2+dx^2+dy^2$, 
and $(\nabla a)^2=\nabla_\mu a \nabla^\mu a$. 

Now, let us define $a=e^\omega$ and suppose $|\omega|\ll 1$. Then, the null-null component of the stress-energy tensor 
becomes 
\begin{align}
& \Exp{T_{\mu\nu}}\hat{l}^\mu \hat{l}^\nu\simeq\nonumber\\
& \frac{1}{4}\Bigl[-(\omega_{,yt}+\omega_{,xy})^2 +\omega_{,yy}(\omega_{,tt}+2\omega_{,xt}+\omega_{,xx})  \Bigr]+O(\omega^3). 
\end{align}
As an example, if one takes $\omega$ as 
\begin{align}
\omega=\epsilon e^{-(t^2+x^2+y^2)}, 
\end{align} 
with $\epsilon\ll 1$, we obtain 
\begin{align}
& \Exp{T_{\mu\nu}}\hat{l}^\mu \hat{l}^\nu \nonumber \\
&=-2\epsilon^2e^{-2(t^2+x^2+y^2)}\{(t+x)^2+2y^2-1\}+O(\epsilon^3). 
\end{align}
It is easily checked that there exist null lines which can violate the ANEC. For instance, consider the curve $y=1$, $x=t$. For this, the l.~h.~s. of Eq.~(\ref{ANEC}) yields a negative value as  
\begin{align}
\int^\infty_{-\infty}\Exp{T_{\mu\nu}}\hat{l}^\mu \hat{l}^\nu d\lambda\simeq -\frac{3\sqrt{\pi}}{2e^2}\epsilon^2<0, 
\end{align}
So, the ANEC is violated. The situation is very similar to the case of conformally coupled scalar field 
in a conformally flat spacetime~\cite{UrbanOlum}.

It is easily checked by the bubble metric~(\ref{bubble-even}) that the bulk spacetime is causally proper 
because the tangent vector $k:=\p_t+ C_1\p_x+ C_2\p_r$ on the bulk causal curve $\gamma$ 
satisfies $C_1\le 1$ and the equality holds only for the boundary null geodesic with $C_2=0$. 

\section{Violation of achronal ANEC in odd-dimensions}
\label{sec:III}
In the previous section, we have shown that the achronal ANEC can be violated in the boundary CFT theory with even dimension, due to 
the conformal anomaly. 
When $d$ is odd, the conformal anomaly terms disappear and 
the stress-energy tensor becomes just the coefficient\cite{HaroSkenderis}
\begin{align}
\label{odd-stress-energy}
& \Exp{T_{\mu\nu}}=\frac{d}{16\pi G_d}g_{(d)\mu\nu}
\end{align}
in the Fefferman-Graham coordinate system~(\ref{Fefferman-Graham}). 
The stress-energy tensor is conformally covariant under the conformal transformation $d\hat{s}_\p^2=a^2ds_\p^2$, just being conformally rescaled with no additive terms. 
Therefore, unless the NEC is violated (i.e., $\Exp{T_{\mu\nu}}$ is negative at some point on the boundary),  
the achronal ANEC cannot be violated (subject to condition~A on the conformal factor). 
In what follows, we set ${d}/{16\pi G_d}=1$, for simplicity. 

We start with the following 6-dimensional 
bubble solutions 
\begin{align}
\label{bubble_six}
& ds^2=\frac{4}{25 r_0^2}\left(r^2-\frac{r_0^5}{r^3} \right)d\chi^2+\frac{dr^2}{r^2-{r_0^5}/{r^3}} \nonumber \\
&+r^2(dx^2+dy^2+dw^2-dt^2) \,. 
\end{align}
where $r\in [r_0,\, \infty]$ and $\chi\in [0, \,2\pi]$.  
The stress-energy tensor~(\ref{odd-stress-energy}) contracted by the null vector $n=\p_t+(5r_0/2)\p_\chi$ becomes 
negative~(see Appendix for $\epsilon=0$), 
\begin{align}
\Exp{T_{\mu\nu}}n^\mu n^\nu=-r_0^2<0. 
\end{align}
Although the ANEC (\ref{ANEC}) is violated along the $S^1$ circle, this does not mean that the achronal ANEC is also violated 
because the closed orbit is not achronal. On the other hand, along the null line with null tangent vector $l=\p_t+\p_x$, 
\begin{align}
\label{zero_NE}
\Exp{T_{\mu\nu}}l^\mu l^\nu=0. 
\end{align}
This suggests that the perturbation of the bubble spacetime~(\ref{bubble_six}) could induce a non-zero stress-energy tensor 
which locally violates the NEC along the achronal null geodesic and hence potentially lead to ANEC violation. We note that one
may desire instead to consider a perturbation of the Poincare AdS solution $(r_0 = 0)$ rather than the bubble solution, since it is
much simpler and Eq.~(\ref{zero_NE}) is still satisfied.  
In that case, however, $pp$-type curvature singularities are generally expected to occur on the horizon 
for generic perturbations~\cite{ChamblinGibbons}. On the other hand, one expects the bubble spacetime to be stable \cite{Horowitz:1998ha}
In the next subsection, we consider the perturbation of the bubble solution~(\ref{bubble_six}). 

\subsection{The perturbed variables}
Let us consider the slightly deformed bubble solution by additing to (\ref{bubble_six}) the following 
static metric perturbations:
\begin{align}
\label{per_metric}
& \delta g_{\mu\nu}=2\epsilon r^2(H\eta_{\mu\nu}S+H_TS_{\mu\nu}), \quad 
\delta g_{\chi \chi }=\epsilon f_{\chi \chi}S \,, 
\end{align}
where $\epsilon$ is a small positive parameter and the Greek indices $\mu, \nu, \dots$ denote the specific choice 
of coordinates $x,\,y,\,w,\,t$ used in (\ref{bubble_six}). Here, $f_{\chi \chi}, \: H, \:H_T$ are functions of $r$, and 
$S$, $S_\mu$, and $S_{\mu \nu}$ are defined in terms of $\rho=\sqrt{x^2+y^2+w^2}$ and a real positive 
parameter $k$ by  
\begin{align}
& S=\frac{\sin k\rho}{\rho} \,, \nonumber \\     
& S_\mu=-\frac{1}{k} D_\mu S, \quad S_{\mu\nu}=\frac{1}{k^2}D_\mu D_\nu S+\frac{1}{4}\eta_{\mu\nu}S \,, 
\end{align}
with the covariant derivative operator $D_\mu$ associated with $\eta_{\mu \nu}$. 
Note that $S$ is regular at $\rho = 0$ and $\lim_{\rho \rightarrow \infty} S =0$, 
guaranteeing the asymptotic convergence of the l.~h.~s. of Eq.~(\ref{ANEC}), as we will show later. 
Note also that $S$ satisfies
\begin{align}
D_\mu D^\mu S+k^2S=0 \,, 
\end{align}  
where one may view the above deformation (\ref{per_metric}) as the Wick-rotated version of a restricted class 
of the scalar-type metric perturbations of the 6-dimensional Schwarzschild-AdS metric. 
Then, following the formulae of \cite{Kodama_Ishibashi2003}, one can derive the equations that determine the three perturbation 
variables $f_{\chi \chi}, H, H_T$:  
\begin{align}
\label{Basic_eq_rM}
&  (r^6-r_0^5\,r)g''(r)+\left(7r^5+\frac{r_0^5}{2}\right)g'(r) \nonumber \\
&\quad\quad\quad\quad-k^2r^2g(r)+4(2r^5+3r_0^5)H'(r)=0, \nonumber \\
& \nonumber \\
& (r^6-r_0^5\,r)H''(r)+2(r^5-r_0^5)H'(r) \nonumber \\
&\quad\quad\quad\quad\quad-\frac{1}{2}(r^5-r_0^5)g'(r)+\frac{k^2r^2}{8}g(r)=0, 
\end{align}
with 
\begin{align}
& H_T(r)=\frac{8}{3k^2r^2}\left(8r^5-3r_0^5 \right)H'(r)-4H(r) \nonumber \\
&\quad\quad\quad+\frac{8}{3k^2r^2}\left(r^5-r_0^5 \right)g'(r)-\frac{2}{3}g(r), 
\end{align}
where $g$ is defined as 
\begin{align}
f_{\chi\chi} = \frac{4}{25r_0^2} \left(r^2-\frac{r_0^5}{r^3} \right)g(r) \,.
\end{align}

Let us expand the functions $H$ and $g$ near $r=\infty$ as 
\begin{align}
\label{parameters_ch}
& H(r)=h_0+\frac{h_1}{r}+\frac{h_2}{r^2}+\frac{h_3}{r^3}+\cdots, \nonumber \\
& g(r)=c_0+\frac{c_1}{r}+\frac{c_2}{r^2}+\frac{c_3}{r^3}+\cdots. 
\end{align} 
Substituting these into (\ref{Basic_eq_rM}), each coefficient is determined by 
\begin{align}
& c_1=h_1=c_3=h_3=0, \nonumber \\
& c_2=-2h_2-\frac{c_0\, k^2}{8}, \nonumber \\
& c_4=\frac{k^2}{32}(16h_2 + k^2c_0),  \quad h_4= -\frac{k^2}{256} (16 h_2 + c_0\,k^2), \nonumber \\
& c_5=-8h_5, \quad\cdots. 
\end{align} 

The regularity condition at the bubble radius $r_0$ is given by~\cite{Mars}
\begin{align}
\label{regularity}
\frac{\nabla_\mu(\xi^2)\nabla^\mu(\xi^2)}{4\xi^2}\to 1, \qquad \xi^2:=g(\p_\chi,\,\p_\chi).  
\end{align}
This implies that $g(r_0)=0$. Note that Eq.~(\ref{Basic_eq_rM}) does not contain $H(r)$, so $h_0$ is a free parameter which does not affect 
the bulk solution. However, achronality along a null geodesic will enforce a relation between $h_0$ and $h_2$, as shown below. 
Hence, there is only one free parameter characterizing the boundary metric at infinity. 
We solve these equations numerically and plot them in Fig.~\ref{numericalsolutions}. 
We also provide an analytic solution for $r_0=0$ in Appendix B which serves as a good approximation for large $k\gg r_0$.

\begin{figure}[t!]
\includegraphics[width=.3\textwidth]{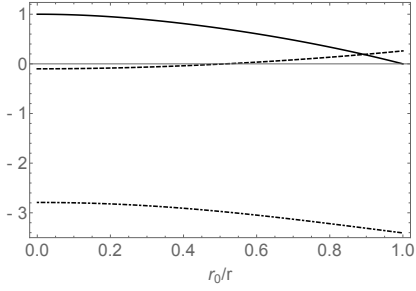}
\caption{\label{numericalsolutions} The functions $g(r)/c_0$ (thick), $H(r)/c_0$ (dashed), $H_T(r)/c_0$ (dot-dashed) for $k= 2r_0$.}
\end{figure}

\subsection{The boundary metric and the stress-energy tensor}
Thanks to the restricted form of our perturbation (\ref{per_metric}), 
one can transform the metric into the Fefferman-Graham form by 
the coordinate transformation: 
\begin{align}
r(z)=\frac{1}{z\left(1 - \frac{r_0^5z^5}{10}\right)}+O(z^{-7}). 
\end{align}
Then, the boundary metric $g_{(0)\mu\nu}$ is given by 
\begin{align}
\label{boundary_metric}
& g_{(0)\mu\nu}dx^\mu dx^\nu= \nonumber \\
& 
\frac{4}{25r_0^2} 
\left(1+\epsilon\, c_0\frac{\sin (k\rho)}{\rho}   \right)d\chi^2 
-\left(1-16\epsilon\, h_2\frac{\sin (k\rho)}{k^2\rho}    \right)dt^2 \nonumber \\
&+\Biggl[1+\frac{8\epsilon}{k^4\rho^3}
\Bigl\{2k(8h_2+k^2h_0)\rho\cos(k\rho) \nonumber \\
&-(k^2h_0(2-k^2\rho^2)+2h_2(8-3k^2\rho^2))\sin(k\rho)\Bigr\}\Biggr]d\rho^2
\nonumber \\
&+\rho^2F(d\theta^2+\sin^2\theta d\phi^2)+O(\epsilon^2),  
\end{align}
where 
\begin{align}
& F=1-\frac{8\epsilon}{k^4\rho^3}\Bigl[-k(8h_2+k^2h_0)\rho\cos(k\rho) \nonumber \\
&\quad\quad\quad-\{k^2h_0+h_2(8-2k^2\rho^2)\}\sin(k\rho)\Bigr]. 
\end{align}
The stress-energy tensor is obtained from the metric coefficient $g_{(5)\mu\nu}$ in the Fefferman-Graham 
expansion~(\ref{Fefferman-Graham}), as explicitly shown in Eq.~(\ref{SE_6}).  As expected, 
the trace of the stress-energy tensor is zero, up to $O(\epsilon)$, i.~e.~, ${{g_{(5)}}^\mu}_\mu=O(\epsilon^2)$ since the trace 
anomaly is zero.   

Now, let us examine the ANEC along a radial null geodesic $l^\mu=(0, l^t, l^\rho, 0, 0)$. Up to $O(\epsilon)$, 
the geodesic equations of motion give
\begin{align}
& l^t=1+\frac{16\epsilon h_2\sin(k\rho)}{k^2\rho}, \nonumber \\
& l^\rho=1+\frac{4\epsilon}{k^4\rho^3}
\Bigl[4h_2k^2\rho^2\sin(k\rho) \nonumber \\
& -(8h_2+h_0\,k^2)\{2k\rho\cos(k\rho)-(2-k^2\rho^2)\sin(k\rho)\} \Bigr].    
\end{align}
Then, the null-null component of the stress-energy tensor is obtained from Eq.~(\ref{SE_6})
\begin{align}
& \Exp{T_{\mu\nu}}l^\mu l^\nu=\epsilon K\times  \nonumber \\
&\quad\quad\quad\frac{2k\rho\cos(k\rho)-(2-k^2\rho^2)\sin(k\rho)}{\rho^3}+O(\epsilon^2),  
\end{align}
where 
\begin{align}
\label{coefficient_K}
K:=\frac{4\{-2k^2h_5+(c_0\,k^2-8h_2)r_0^5\}}{3k^4}. 
\end{align}
As expected, the NEC is locally violated unless $K=0$. On the other hand, the ANEC~(\ref{ANEC}) is 
satisfied, up to $O(\epsilon)$ because 
\begin{align}
\int^\infty_{-\infty}\Exp{T_{\mu\nu}}l^\mu l^\nu d\lambda=O(\epsilon^2), 
\end{align}
where $\lambda$ is the affine parameter of $l^\mu$. 

For the boundary metric $g^{(0)}_{\mu\nu}$, the higher order corrections in $\epsilon$ can be set to zero by 
the following argument. Let us denote the bulk metric by ${\Phi}$~\footnote{Here, we omit the indices for simplicity.}. It can be 
expanded by 
\begin{align}
{\Phi}={\Phi}_0+\epsilon {\Phi}_1+\epsilon^2{\Phi}_2+\epsilon^3{\Phi}_3+\cdots. 
\end{align}
The equations of motion for the variables are obtained from the vacuum Einstein equations as 
\begin{align}
& {\cal L}{\Phi}_1=0, \nonumber \\
& {\cal L}{\Phi}_i={S}_i({\Phi}_1,\,{\Phi}_2, \cdots, {\Phi}_{i-1}), \quad i=2,3,\cdots, 
\end{align}
where ${\cal L}$ is a linear second order differential operator, and ${S}_i$ is a function of the variables 
${\Phi}_j, \,\, j=1,2,\cdots, i-1$. One can formally construct the solution ${\Phi}_i$ as ${\Phi}_i=\int {G} {S}_i d^6x$ 
in terms of the Green function ${G}$ of the linear operator ${\cal L}$ satisfying ${\cal L}{G}=\delta(x-x')$. 
The boundary conditions of ${ G}$ are the regularity condition at the bubble radius and the normalization condition 
at infinity. So, the perturbed boundary metric $\delta g^{(0)}_{\mu\nu}$ for higher corrections in $\epsilon$ can be always set to zero. 

\begin{figure}[t!]
\includegraphics[scale=.3]{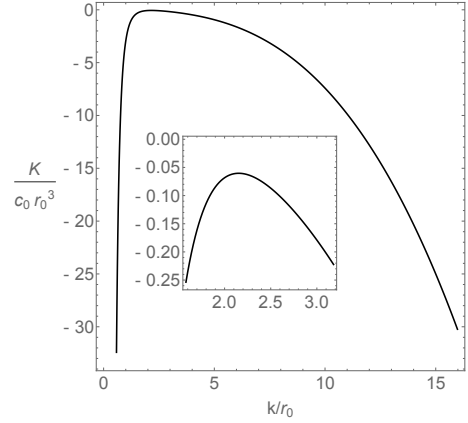}
\caption{\label{Kfigure} The function $K$ in the bubble spacetime. Notably, this never vanishes, as shown in the inset.}
\end{figure}

We can check that the achronality along the null geodesic curve of $l^\mu$ follows from choosing the parameter $h_0$ in Eq.~(\ref{parameters_ch}) as 
\begin{align}
\label{condition2}
h_0=-8\frac{h_2}{k^2}. 
\end{align}
In this case, the boundary metric reduces to 
\begin{align}
& g_{(0)\mu\nu}dx^\mu dx^\nu=\frac{4}{25r_0^2}\left(1+\epsilon\, c_0\frac{\sin (k\rho)}{\rho}   \right)d\chi^2 +\nonumber \\
&\left(1-16\epsilon\, h_2\frac{\sin (k\rho)}{k^2\rho}    \right)[-dt^2+d\rho^2+\rho^2(d\theta^2+\sin^2\theta d\phi^2)]. 
\end{align}
Note that $\chi=\mbox{const.}$ hypersurface is conformally flat. This means that any two points along the null curve 
with the tangent vector $l^\mu$ cannot be connected by a timelike curve within the $\chi=\mbox{const.}$ hypersurface. Since 
$\p_\chi$ is a spacelike Killing vector orthonomal to $l^\mu$, the null geodesic curve is also the fastest causal curve among the causal curves 
with the tangent vector $l'^\mu=l^\mu+A({\p_t})^\mu+B (\p_\chi)^\mu$ with $A$, $B$ some 
constants ($A$ is non-negative and can vanish only when $B=0$), thereby guaranteeing the achronality of the null geodesic curve. For $k\neq 0$, 
one can always enforce the condition~(\ref{condition2}). Furthermore, in Fig.~\ref{Kfigure}, we show that
(\ref{coefficient_K}) never vanishes for $c_0\neq 0$. This implies that the achronal ANEC can be violated 
after a conformal transformation, as shown below.

\subsection{Conformal transformation}
In the Fefferman-Graham form, the perturbed metric~(\ref{per_metric}) is written by 
\begin{align}
\label{FG_six}
ds_6^2=\frac{dz^2+g_{\mu\nu}(\rho, z)dx^\mu dx^\nu}{z^2}. 
\end{align} 
As done in Sec.~II, one can transform the metric into a different Fefferman-Graham form 
\begin{align}
\label{FG_six_II}
\hat{ds}_6^2=\frac{dZ^2+\hat{g}_{\mu\nu}(\xi, Z)d\hat{x}^\mu d\hat{x}^\nu}{Z^2}, \quad \hat{x}^\mu=\xi,\,t,\,\theta,\,\varphi
\end{align}
with a boundary metric
\begin{align}
\label{conformal_transformation_six}
\hat{g}_{\mu\nu}(\xi, 0)d\hat{x}^\mu d\hat{x}^\nu=a^2(\xi)g_{\mu\nu}(\rho, 0)dx^\mu dx^\nu, 
\end{align} 
for an arbitrary scale factor $a(\xi)$. 
 
As the coordinate transformation, we make an ansatz;   
\begin{align}
& z(\xi,\,Z)=Z\left(\frac{1}{a(\xi)}+\alpha_1(\xi)Z+\alpha_2(\xi)Z^2+\cdots   \right), \nonumber \\
& \rho(\xi,\,Z)=\xi+\beta_1(\xi)Z+\beta_2(\xi)Z^2+\beta_3(\xi)Z^3+\cdots.  
\end{align}
By substituting these into Eq.~(\ref{FG_six}) and keeping the metric form~(\ref{FG_six_II}), each coefficient is determined by  
\begin{align}
& \alpha_{2i+1}=\beta_{2i+1}=0 \quad \mbox{for} \quad i\in 0,1,2,\cdots, \nonumber \\
& \alpha_2=-\frac{a'^2}{4a^5R_0}, \qquad \beta_2=\frac{a'}{2a^3R_0}, \nonumber \\
& \alpha_4=\frac{2a^2a'^2R_2+a'^4}{16a^9R_0^2}, \nonumber \\
& \beta_4=-\frac{a'}{16a^7R_0}\{2R_0(2a^2R_2+a'^2)+aa'R_0' \}, \cdots, 
\end{align}
where $R_i$ is the coefficient in the expansion of $g_{\rho\rho}$, 
\begin{align}
g_{\rho\rho}=R_0(\xi)+R_2(\xi)z^2+R_4(\xi)z^4+\cdots. 
\end{align}

Under the coordinate transformation, the boundary metric satisfies Eq.~(\ref{conformal_transformation_six}) 
and the stress-energy tensor are transformed into
\begin{align}
\hat{T}_{\mu\nu}=\frac{1}{a^3(\xi)}T_{\mu\nu}. 
\end{align} 
Since this is the conformal transformation on the boundary metric, the achronal null geodesic orbit does not change and the 
tangent vector is transformed by
\begin{align}
\hat{l}^\mu=\frac{1}{a^2(\xi)}l^\mu. 
\end{align}
Hence, the averaged null energy condition is transformed into 
\begin{align}
\int^\infty_{-\infty}\Exp{\hat{T}_{\mu\nu}}\hat{l}^\mu\hat{l}^\nu d\hat{\lambda}=\int^\infty_{-\infty}\frac{1}{a^5}\Exp{T_{\mu\nu}}l^\mu l^\nu d\lambda, 
\end{align}
where we used the fact that $d\hat{\lambda}=a^2d\lambda$ for the affine parameter $\hat{\lambda}$ 
for the null geodesic in $(\partial M, \hat{g})$. 
This implies that for a suitable choice of the scale factor $a$, this becomes negative unless $K=0$ in Eq.~(\ref{coefficient_K}). As an explicit simple example, choose
\begin{align}
a(\xi) = (1+b^2 e^{-k^2\xi^2})^{-1/5}
\end{align}
for real $b$ which gives
\begin{align}
&\int_{-\infty}^\infty \frac{1}{a^5}\Exp{T_{\mu\nu}}l^\mu l^\nu d\lambda \nonumber\\
&\quad\quad=-2k^2 b^2 K \left(\frac{\sqrt{\pi }}{\sqrt[4]{e}}-\pi  \text{Erf}\left(\frac{1}{2}\right)\right)\epsilon +O(\epsilon^2) < 0.
\end{align}

\subsection{The generic condition in the bulk}
Compared with the bubble solution with high symmetry in Sec.~II, it is not immediate obvious whether or not 
the perturbed bubble solution is causally proper. 
However, as shown in the Gao-Wald theorem~\cite{GaoWald}, the condition of being causally proper
is always satisfied provided that the following three conditions are satisfied in the bulk: 
\begin{enumerate}
\item the NEC holds for the bulk null geodesics, 
\item no causal pathologies are observable from the boundary, e.g. singularities or regions of causality violation,  
\item the null generic condition holds for the bulk null geodesics~\footnote{
The null generic condition is the statement that any null geodesic with the tangent $l^a$ contains a point where 
$l_{[a}R_{b]cd[e}l_{f]}l^cl^d \neq 0$. For the vacuum case, the Riemann tensor can be replaced with the Weyl tensor.}
\end{enumerate}
The only non-trivial check is the final condition 3, as the perturbed vacuum spacetime automatically 
satisfies the first and the second conditions. If there existed a bulk null geodesic orbit with tangent vector $k^a$ 
that connects two points on the boundary achronal null geodesic with the tangent vector $l^\mu$, the orbit of $k^a$ would 
have to be sufficiently near the conformal boundary, and hence, $k^z$ would be small, i.~e., $k^z=O(\epsilon)$ because we consider only 
perturbations of the bubble spacetime~(\ref{bubble_six}) which is causally proper. 
The Weyl tensor in the bulk behaves as  $C_{cadb}k^a k^b\sim C_{a\mu b\nu}l^\mu l^\nu=O(\epsilon)$. 
For example, 
\begin{align}
C_{\chi\mu \chi\nu}l^\mu l^\nu\sim \epsilon\frac{2k\rho\cos(k\rho)-(2-k^2\rho^2)\sin(k\rho)}{\rho^3}. 
\end{align} 
Then, the null geodesic would necessarily pass through a point in which the generic condition is satisfied. 
This is impossible by the theorem~\cite{GaoWald}. 
\section{Time delay and the weak cosmic censorship in AdS}
\label{sec:IV}

In the previous section, we have shown within perturbative framework that the spacetime considered is 
causally proper from the Gao-Wald theorem~\cite{GaoWald}. We can also deduce from similar arguments to those in the 
Gao-Wald theorem the following Proposition concerning weak cosmic censorship in asymptotically AdS spacetimes.  

\medskip

\noindent  
{\it Proposition. } \\
Suppose $(M,\,g_{ab})$ is an asymptotically AdS spacetime, which can be conformally embedded in an unphysical spacetime $(\tilde{M},\,\tilde{g}_{ab})$ so that with a smooth function $\Omega$ in $\tilde{M}$, we have $\tilde{g}_{ab}=\Omega^2g_{ab}$ and $\Omega=0$ on the timelike boundary $\partial M$ in $\tilde M$. 
Suppose $(M,\,g_{ab})$ satisfies the following conditions,
\begin{itemize}
\item[(i)] the NEC and the null generic condition, 
\item[(ii)] $\bar{M}:=M\cup \p M$ is strongly causal, and $\partial M$ itself is globally hyperbolic. 
\end{itemize} 
If there is a causal curve in $\bar M$ from a point $p\in \p M$ connecting to a point 
$q\in E^+(p,\,\p M) \setminus \{ p \}$ which passes through points in the bulk $M$ (i.e., which is not entirely in $\partial M$), 
then there must be a past-incomplete null geodesic curve in $M$ from a point of $\p M$.
That is, there is a singularity in $M$ visible from a boundary point in the future of $p$, implying 
a violation of weak cosmic censorship.  

\medskip 

\noindent 
{\it Proof. } \\
Let us consider $p \in \partial M$ and $E^+(p, \partial M)$. Let $q$ be a point in $E^+(p, \partial M) \setminus 
\{ p\}$ and $\lambda$ be a null geodesic generator of $E^+(p, \partial M)$ that connects $p$ and $q$. 
Then, by assumption, there exists a future-directed causal curve $\mu$ in $\bar M$ from $p$ to the future end point $q$, 
passing through the bulk $M$. 
Now suppose that $\mu$ is a null geodesic generator of $\dot{J}^+(p, \bar{M})$.  
Then, $\mu$ must be an achronal null geodesic and also be complete as it connects the two points $p,q$ at infinity ($\partial M$). 
However, if $\mu$ is a complete null geodesic in $M$, it would admit a pair of conjugate points 
due to the condition (i), and hence fail to be an achronal null geodesic, according to the proposition 4.5.12 in~\cite{HawkingEllis}. 
Thus, $\mu$ cannot be a null geodesic generator of $\dot{J}^+(p, \bar{M})$. By the proposition 4.5.10 in~\cite{HawkingEllis}, 
$p$ and $q$ can be joined by a timelike curve, implying in particular that the future end point $q$ of $\mu$ cannot be in $E^+(p, \bar{M})$. 
Then, since $\lambda\cap E^+(p, \bar{M})\setminus \{ p\}\neq \emptyset$ by the condition (ii), 
there must be a future end point $r$ of $\lambda\cap E^+(p, \bar{M})$ such that $r\in J^-(q,\partial M)\setminus \{ q\}$ and $r$ is intersected by a bulk null geodesic generator 
$\gamma$ of $\dot{J}^+(p, \bar{M})$, which entirely lies in $M$ except their end points on $\partial M$. 

If $\gamma$ had a past endpoint, $\gamma$ would be a generator of $E^+(p,\,\bar{M})$ with the past end point $p$. 
But, this is a contradiction, since there would be a 
pair conjugate points along $\gamma$ between $p$ and $r$, and hence $\gamma$ could not be a 
generator of $E^+(p, \bar{M})$, again by the proposition 4.5.12 in~\cite{HawkingEllis}. 
Thus, $\gamma$ has no past end point. If $\gamma$ were past-complete 
inextendible null geodesic, there would be a pair conjugate points along $\gamma$ which also leads to contradiction. So, $\gamma$ 
must terminate at a singularity in the past direction. Since $\gamma$ is future directed from $p \in \partial M$, this singularity is visible from $q$, that is, nakedly singular.  \hfill $\Box$

\medskip 

\begin{figure}[t!]
  \includegraphics[scale=.5]{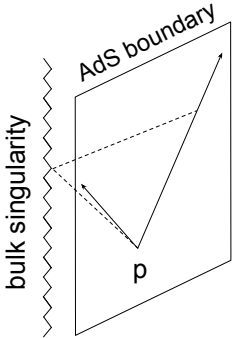}
  \caption{\label{gaowald}The solid and dashed lines represent $\lambda\in E^+(p, \bar{M})$ and 
$\gamma\in \dot{J}^+(p, \bar{M})$, respectively. $\gamma$ terminates at the singularity.}
\end{figure}

It may be instructive to see a concrete example. For the negative mass planar Schwarzschild-AdS spacetime, 
$E^+(p, \p M)$ is intersected by a bulk null geodesic generator~(dashed curve) of $\dot{J}^+(p, \bar{M})$ 
which terminates at singularity, as shown in Fig.~\ref{gaowald}.

\section{Summary and discussions} 
\label{sec:V} 
In this paper, we have explored possible interplay between the archonal ANEC, the causally proper nature of bulk and boundary spacetimes, 
and the weak cosmic censorship within the context of AdS/CFT duality. We have shown that the achronal ANEC can be violated in 
holographic theories by vacuum bubble AdS solutions. 
In Sec.~\ref{sec:II}, we have shown a violation of the achronal ANEC in $4$-dimensional 
boundary CFT due to the conformal anomaly. Since the boundary spacetime is a curved spacetime, 
the violation does not conflict with the proof of the ANEC in the flat boundary 
spacetime~\cite{KellyWall2014}. As shown in the Gao-Wald theorem~\cite{GaoWald}, our 
examples are ``causally proper" in the sense that a ``fastest null geodesic" connecting 
any two points on the boundary must lie entirely on the boundary. This means that there is no 
acausal signaling in the boundary theory, which is not physically permitted.    

In Sec. \ref{sec:III}, we have shown the violation of the achronal ANEC in $5$-dimensional CFT 
by perturbing the $6$-dimensional vacuum bubble spacetime. In this case, since the boundary is $5$-dimensional, there is no
conformal anomaly. 
The extent of the violation is small since 
the null-null component of the boundary stress-energy tensor is proportional to the amplitude of the 
perturbation. So, it would be interesting to investigate if the ANEC 
is also violated beyond the perturbation. One of the candidates is the vacuum bubble solution 
with a wormhole geometry in the boundary spacetime. In the thermal state, the vacuum black hole 
solution with a wormhole throat on the boundary has been numerically constructed~\cite{IMM2018}.  
Even though the ANEC is not violated by the existence of the infinite thermal energy in the asymptotic flat 
region, negative null energy appears near the throat, caused by the negative curvature on the horizon. 
This leads us to speculate that the ANEC would be violated for a vacuum bubble AdS solution with a wormhole 
throat, as there is no asymptotic positive null energy.     

In Sec. \ref{sec:IV}, we have presented a Proposition which connects the cosmic censorship in the AdS bulk to the 
causally proper nature of our holographic system. According to the proposition, acausal propagation of signals 
in the boundary theory means the occurrence of a naked singularity in the bulk. 
In \cite{IshibashiMaeda12}, Hawking and Penrose type singularity theorems \cite{HawkingEllis} have been revisited in asymptotically AdS spacetimes, 
and the essential role of the strong gravity condition (i.e., the existence of a {\em trapped set}) in the bulk has been discussed.  
The above proposition may be viewed as a different type of singularity theorem which does not need to impose 
the strong gravity condition in the bulk, but which, instead, invokes the causally proper nature, as an alternative condition 
that involves sensible causal interactions between the bulk and the conformal boundary.  
This may give some new insights into possible applications of the AdS/CFT duality, in particular new connections between 
bulk and boundary causality.  

\section{Acknowledgements}
We would like to thank T.~Okamura for useful discussions. This work was supported in part by JSPS KAKENHI Grant Number 
17K05451~(KM), 15K05092~(AI) and the European Research Council (ERC) under the European Unions Horizon 2020 research and 
innovation programme (grant agreement No. 758759)~(EM). 

\medskip 

\appendix 
\section{Stress energy tensor for $d=5$-dimensions}
For general $g_0,\, h_0$, the stress tensor is
\begin{widetext}
\begin{align}
\label{SE_6}
& g^{(5)}_{\mu\nu}dx^\mu dx^\nu=-\frac{16r_0^3}{125}\left[1+\epsilon\left(10\frac{h_5}{r_0^5}+c_0\right)\frac{\sin(k\rho)}{\rho} \right]d\chi^2
+\frac{r_0^5}{5}\left[-1+\epsilon\left\{8h_2+5k^2\left(-8\frac{h_5}{r_0^5}+c_0 \right)\right\}\frac{\sin(k\rho)}{3k^2\rho}\right]dt^2
\nonumber \\
&+\frac{r_0^5}{5}\Biggl[1+\frac{\epsilon}{3k^4\rho^3}\Biggl\{8k\Bigl(-\frac{10k^2h_5}{r_0^5}+8h_2+k^2(5c_0+6h_0)\Bigr)\rho\cos(k\rho)                 
\nonumber \\ 
&\quad\quad\quad\quad+\Bigl\{-64h_2+k^2\Bigl(\frac{80h_5}{r_0^5}-8(5c_0+6h_0)+3(8h_2+k^2(5c_0+8h_0)\rho^2)\Bigr)\Bigr\}\sin(k\rho) \Biggr\}\Biggr]d\rho^2
\nonumber \\
&+\frac{r_0^5}{5}\Biggl[\rho^2+\frac{4\epsilon}{3k^3}\Bigl\{\frac{10k^2h_5}{r_0^5}-8h_2-k^2(5c_0+6h_0)\Bigr\}\cos(k\rho)         
\nonumber \\
&\quad\quad\quad\quad+\frac{\epsilon}{3k^4}\Bigl\{32h_2+k^2\Bigl(-\frac{40h_5}{r_0^5}(1-k^2\rho^2)
+20c_0+24h_0-8h_2\,\rho^2-5c_0k^2\,\rho^2\Bigr)\Bigr\}\frac{\sin(k\rho)}{\rho}\Biggr](d\theta^2+\sin^2\theta d\phi^2)+O(\epsilon^2). 
\end{align}
\end{widetext}
\section{Analytic solution for $r_0=0$}
Despite being unstable to generic perturbations, it is useful to note that Eq.~({\ref{Basic_eq_rM}}) can be solved analytically when $r_0=0$. The solutions are
\begin{align}
g(r) &= \frac{c_0}{3} \frac{e^{-k/r}(k^2+3kr + 3r^2)}{r^2}\nonumber\\
H(r) &= h_0 + \frac{c_0-g(r)}{8}\nonumber\\ 
&= -\frac{c_0}{24}\left(1+ \frac{e^{-k/r}(k^2+3kr + 3r^2)}{r^2}\right)\nonumber\\
H_T(r) & = -4h_0 -\frac{3c_0+g(r)}{6}\nonumber\\
&=\frac{c_0}{18}\left(3- \frac{e^{-k/r}(k^2+3kr + 3r^2)}{r^2}\right)
\end{align}
The second equalities are for the choice $h_0 = -8h_2/k^2$. For any $h_0$, this solution leads to 
\begin{align}
K = - \frac{k^3}{135}c_0.
\end{align}
At large $k$, the solution for $r_0\neq 0$ is approximately equal to the vacuum solution confirming that $K$ never vanishes in the perturbed bubble spacetime. The boundary metric is (to all orders in $\epsilon$)
\begin{align}
&g_{\mu\nu}^{(0)}dx^\mu dx^\nu = \frac{4}{25r_0^2}\left(1+\epsilon c_0\frac{\sin(k\rho)}{\rho}\right)d\chi^2+\nonumber\\
&\left(1-\epsilon \frac{c_0}{3} \frac{\sin(k\rho)}{\rho}\right)[-dt^2+d\rho^2+\rho^2(d\theta^2+\sin^2\theta d\phi^2)].
\end{align}
and
\begin{align}
l^t = l^\rho = \left[1-\epsilon \frac{c_0}{3}\frac{\sin(k\rho)}{\rho}\right]^{-1}.
\end{align}
For our previous choice of conformal factor $a(\xi) = (1+b^2\exp(-k^2\xi^2))^{-1/5}$, we find that
\begin{align}
&\int_{-\infty}^\infty\Exp{T_{\mu\nu}}l^\mu l^\nu d\lambda \nonumber\\
&\quad\quad\approx -\epsilon (3.78\times 10^{-3})k^5 b^2  +O(\epsilon^2) < 0.
\end{align}

\end{document}